\documentstyle[12pt,epsf]{article}

\input{psfig}

\newcommand{\nc}{\newcommand}
\def\frac#1#2{{\textstyle {#1 \over #2}}}

\nc{\beq}{\begin{equation}}
\nc{\eeq}{\end{equation}}
\nc{\beqa}{\begin{eqnarray}}
\nc{\eeqa}{\end{eqnarray}}
\nc{\lsim}{\begin{array}{c}\,\sim\vspace{-21pt}\\< \end{array}}
\nc{\gsim}{\begin{array}{c}\sim\vspace{-21pt}\\> \end{array}}
\nc{\eps}{\epsilon}
\nc{\s}{\sigma}
\nc{\veps}{\varepsilon}
\nc{\no}{\noindent}
\nc{\D}{\Delta}

\begin{document}
\begin{titlepage}

\begin{center} 

\vskip .5 in
{\large \bf 
Critical Behavior of Coupled q-state Potts Models under Weak Disorder}

\vskip .6 in

  {\bf  P. Simon}\footnote{simon@lpthe.jussieu.fr }
   \vskip 0.3 cm
 {\it   Laboratoire de Physique Th\'eorique et Hautes Energies  }\footnote{ 
 Unit\'e associ\'ee au CNRS URA 280}\\
 {\it  Universit\'es Pierre et Marie Curie Paris VI et Denis Diderot Paris 
VII}\\
{\it  2 pl. Jussieu, 75251 Paris cedex 05 }

  \vskip  1cm   
\end{center}
\vskip .5 in

 \begin{abstract}
We investigate the effect of weak disorder on different coupled $q$-state Potts 
models with
$q\le 4$ using two loops renormalisation group. This study presents new  
examples of first order transitions driven by randomness. We found that weak 
disorder makes the models decouple.  Therefore, it appears that no relations 
emerge, at a perturbation level, between the disordered $q_1\times q_2$-state  Potts 
model and the two  disordered $q_1$, $q_2$-state Potts models ($q_1\ne q_2$), 
despite their central charges are similar according to recent numerical 
investigations. Nevertheless, when two $q$-state Potts models are considered 
($q>2$), the system remains always driven in a strong coupling regime,
violating apparently the Imry-Wortis argument.

 \end{abstract}

PACS NUMBERS: 05.70.Jk; 64.60.Fr; 75.10.Hk; 75.40.Cx

\end{titlepage}

\renewcommand{\thepage}{\arabic{page}}
\setcounter{page}{1}
\noindent
 The effect of weak randomness on continuous phase transitions has been much 
studied either analytically \cite{ludwig,dots}, or numerically \cite{num}.  For 
most of these cases, the Harris criterion \cite{harris} provides us a good 
method to see if disorder will change the universality class of the transition. 
However, it has been realized recently, that a weak bond randomness can also 
have strong effects on first order transitions $2D$ systems and can induce a 
second order
phase transition in a system that would undergo a first order one \cite{hui}. 
This result  has been established more rigorously in \cite{aiz}.
This has been first  tested with success on systems presenting a fluctuation 
driven first order transition, namely transitions which are expected to be 
continuous at a mean field level, but become first order when transitions are 
incorporated. Hence, Cardy has shown that the addition of weak disorder on a 
system of $N$ coupled Ising models ( a model presenting a first order transition 
driven by fluctuations) makes the system flow to $N$ decoupled Ising models 
\cite{cardy}. This study has then been extended to the case of $N$ $3-$state 
Potts models by Pujol; the result was a non-Ising like second order transition 
(for $N>2$) \cite{pujol}. The $q-$states Potts model with $q>4$
appears  as a more natural model to test and analyse more deeply the effects of 
disorder,
because the first order transition is now of mean field type, so much stronger. 
Nevertheless, there is no analytical approach able to control the effect of 
disorder on such systems.  On the other hand, Chen
{\it et al.} have investigated the $8$-states potts Model using Monte Carlo 
simulations and confirmed the transition to be continuous but also found 
numerical values of the critical exponents consistent with those of the pure 
Ising model \cite{chen}. Yet, more recent  numerical studies of Cardy 
and Jacobsen \cite{cardy1} and of Picco \cite{picco} are in a clear disagreement 
with the latter conclusion. They found a magnetic exponent $\beta/\nu$ which 
varies continuously with $q$. It has also been shown, using finite size scaling 
combined with conformal invariance, that the values of the central charges are 
related to one another by a factorization law $c(q=q_1\times q_2)=c(q_1)+c(q_2)$. 
Therefore, this measure
is not able to distinguish between a non trivial random behavior for a 
$q=q_1\times q_2$ state
random Potts model and two decoupled $q_1$ and $q_2$ state Potts models. This 
could suggest some links between a disordered $q=q_1\times q_2$-state Potts model and 
the corresponding disordered $q_1,q_2$-state potts models despite the magnetic 
exponents appear different for the $4,8$-state Potts models \cite{cardy1,picco}. 
\\Up to now, the only analytical results concern several $q$-state Potts models 
with disorder. We intend in this letter to analyse  the case of two different 
Potts models with disorder.
Therefore, we investigate analytically (in a perturbative sheme in  powers of 
$(q-2)$) the behavior of two coupled $q_1$ and $q_2$-state potts models with 
$q_1,~ q_2 \in \{2,3,4\}$ in order to compare it to the numerical results of the 
$q=q_1\times q_2$-state Potts model. We find a rather complex situation: 
When two different models are considered, there is a factorisation law and the 
models decouple. Therefore, our analysis confirms no apparent relations (at 
least perturbatively!)
between the $q=q_1\times q_2$-state Potts model and the corresponding disordered 
$q_1,q_2$ state Potts models,
despite their central charge to be similar. Nevertheless, when $q_1=q_2=3$ or $4$, 
the system 
flows in a strong coupling regime, despite the presence of disorder, what is in 
apparent contradiction with the Imry-Wortis argument.\\

Our model consists of one $q_1$-state Potts  model coupled to a $q_2-$state 
Potts model by their energy operators ($q_i\in \{2,3,4\})$. The Hamiltonian of 
the system has the following form
\beq
\label{ham1}
H=H_{1}+H_{2}-g_{12}\int d^2 x~\veps_1\veps_2
+ m_1\int d^2 x~\veps_1+ m_2\int d^2 x~\veps_2~.
\eeq
\no
$H_1,~H_2$ are respectively the Hamiltonians of the pure $q_1$ and $q_2$ state 
Potts models, $m_i$ the reduced temperatures ($i=1,2$), $\veps_i$ corresponds to 
the energy operators of the pure models and finally $g_{12}$ is the coupling 
constant associated to the interaction term.
 The partition function can hence be written as
\beq
Z=Tr_{s_{1,i}}Tr_{s_{2,i}}e^{[S_1^0+S_2^0+g_{12}\int d^2 x~\veps_1\veps_2
+ m_1\int d^2 x~\veps_1+ m_2\int d^2 x~\veps_2]}
\eeq
where $s_{1,i}$ is a spin operator of the $q_1$-state Potts model. 
Therefore, a correlation
function $<O(0)O(R)>$, where $O$ is some local operator, is expanded 
perturbatively like:
$$
<O(0)O(R)> = <O(0)O(R)>_0+<S_IO(0)O(R)>_0+{1\over2}<S_I^2O(0)O(R)>_0+\cdots
$$
where $<>_0$ means the expectation value taken with respect to $S_1^0+S_2^0$ and 
$S_I =  \int  d^2 x~\veps_1 \veps_2$.
The calculation of correlation functions can be performed with the Coulomb-gas 
representation \cite{dots1}. The central charge of a $q_i$-state Potts model is 
written as $c={1\over 
2}+\epsilon_i$, where $\epsilon_i$ can also be used as a short distance 
regulator for 
the integrals involved in correlation functions calculations. Then the limit 
$\eps\to 0$ corresponds to the Ising model while the Potts model is obtained 
for some finite value of $\eps$. The recent numerical results of Cardy {\it et 
al.} \cite{cardy1} have
proved that the $(q-2)$ expansion gives accurate results for $q=3$ and good 
qualitative results for $q=4$ (with an error around $4\%$ on the magnetic 
exponent  of the disordered $4-$state potts model).\\
When we have only one coupling constant $g_0$, 
its renormalisation is determined
directly by a perturbative computation. $g$ is also
given by the Operator Algebra (O.A.) producing
\beq
\label{exrg}
g_0 \int \displaystyle\varepsilon_a(z)\varepsilon_b(z) d^2z
+ {1\over 2} \left(g_0 \int \displaystyle
\varepsilon_a(z)\varepsilon_b(z) d^2z\right)^2 +\cdots= g\int 
\displaystyle\varepsilon_a(z)\varepsilon_b(z) d^2z~,
\eeq
with $g=g_0 + A_2 g_0^2  + \cdots $ where $A_2$ comes from the contraction
\beq
{1\over 2} \int \displaystyle 
\varepsilon_a(z)\varepsilon_b(z) d^2z \int \displaystyle 
\varepsilon_c(z)\varepsilon_d(z) d^2z = A_2 \int
\displaystyle\  \varepsilon_a(z)\varepsilon_b(z) d^2z 
\eeq
\noindent
 Therefore, in  the limit $m_i\to 0$, the $2$-loop renormalisation group 
equation associated to (\ref{ham1}) reads
\beq
\label{beta1}
\beta(g_{12}) \equiv {dg_{12}\over d\log r}= {\eps_1+\eps_2\over 2} 
g_{12}(r)+o(g_{12}^3)
\eeq
\no
Here, $\eps_i=2-\Delta_{\veps_i}$, with $ \Delta_{\veps_i}$ the physical 
dimension of the
energy operator $\veps_i$. For more details, we refer to references 
\cite{dots1,dots}. In our notation, we have $\eps=0$ for the Ising model, 
$\eps={2\over 5}$
for the $3-$state Potts model and  $\eps=1$ for the $4-$state Potts model.  From 
the equation (\ref{beta1}), we clearly see that the system is driven
in a strong coupling regime indicating probably a first order transition. A 
similar situation has already been encountered,
when one couples two $q$-state Potts models \cite{pujol}.  Moreover, it can be 
proved exactly that
we have a mass gap generation \cite{vays}. The equation  (\ref{beta1}) can be 
generalized
for one Ising model coupled to several Potts models giving a new interesting 
fixed point structure
\cite{simon}.  \\
We now add quenched randomness coupled to the local energy densities. This can 
be done by
introducing in (\ref{ham1})  position dependent random mass terms $m_i \to 
m_i(x)$
with $\overline{m_i(x)}=0$ and $\overline{m_i(x)m_j(y)}=\Delta_{ij}\delta(x-y)$. 
$\Delta_{ij}$ represents the $2\times 2$ symmetric covariance matrix whose 
elements are strictly positive as it should be. If we have considered a diagonal 
covariance matrix corresponding to  independent disorders for each models, then 
$\Delta_{12}$ would have been generated by the R.G. equations. We then apply the 
replicated method by introducing $n$ copies of the system and averaging Gaussian 
distributions for $m_i(x)$. We finally obtain the sum
of $n$ $q_1$ and $q_2$-state Potts models coupled by their energy densities. The
Hamiltonian can be written as
\beq
\label{ham2}
H=H_{1}+H_{2}-g_{12}\int d^2 x~\sum\limits_a\veps_1^a\veps_2^a-\int d^2 
x~\sum\limits_{<i,j,a,b>}\Delta_{ij}\veps_i^a\veps_j^b~,
\eeq
where $i,j=1,2$ and $a,b$ runs from $1$ to $n$ and $<\dots>$ means that when 
$i=j$ then $a\ne b$. The hamiltonian (\ref{ham2}) has four coupling constants. 
It is a generalization of the study of two Ising models with disorder 
\cite{cardy} or two
Potts models with disorder \cite{pujol}. We now derive the $2-$loops R.G. 
equations associated to (\ref{ham2}). The procedure employed in a generalisation 
of (\ref{exrg}). The methods we use are explained and detailed in ref.
\cite{dots}. Nevertheless, since we have to compute correlations functions 
involving
different energy operators, new integrals appear at one and two loops. Details 
of the algebra will be
presented elsewhere \cite{simon1}. Therefore, taking the replica limit $n\to 0$, 
the four beta functions are, at two loops

\beqa
\label{rg}
\beta_{g_{12}}\equiv {dg_{12}\over d\log l}&=&{\eps_1+\eps_2\over 2}g_{12} - 
g_{12}(\D_{11}+\D_{22})+{1\over 2} 
g_{12}(\D_1^2+\D_2^2)+3\D_{12}^2g_{12}+\D_{12}g_{12}^2\nonumber\\
\beta_{\D_{11}}\equiv {d\D_{11}\over d\log 
l}&=&\eps_1\D_{11}-2\D_{11}^2+2g_{12}\D_{12}+2\D_{11}^3-2\D_{11}\D_{12}g_{12}-\D
_{11}g_{12}^2\nonumber\\
\beta_{\D_{22}}\equiv {d\D_{22}\over d\log 
l}&=&\eps_2\D_{22}-2\D_{22}^2+2g_{12}\D_{12}+2\D_{22}^3-2\D_{22}\D_{12}g_{12}-\D
_{22}g_{12}^2\\
\beta_{\D_{12}}\equiv {d\D_{12}\over d\log l}&=&{\eps_1+\eps_2\over 
2}\D_{12}+g_{12}(\D_{11}+\D_{22})-\D_{12}(\D_{11}+\D_{22})\nonumber\\
&&+\D_{12}\left[{1\over 
2}(\D_{11}^2+\D_{22}^2)+\D_{12}^2-2\D_{12}g_{12}-g_{12}^2\right]\nonumber
\eeqa
In these equations we have made the change $g_{12}\to 4 \pi g_{12}$ and $\D\to 4 
\pi \D$.
We have not considered replica symmetry breaking. Moreover, recent analytical 
and numerical results on disordered Potts models are in favor of a non replica 
symmetry breaking  scenario \cite{dots2}.
 When we couple two similar models (Ising or Potts), we have $\eps_1=\eps_2$ and 
$g_{12}=g$,
$\Delta_{11}=\Delta_{12}=\Delta_{22}=\Delta$ and the flow reduces to
\beqa
\label{rg2}
{dg \over d\log l} &=& \epsilon g - 2g\Delta + g^2 \Delta
+ 4g \Delta^2 \nonumber\\
{d\Delta \over d\log l} &=& \epsilon \Delta -2 \Delta^2 + 2 \Delta g +2 \Delta^3
-  g^2 \Delta - 2\Delta^2 g
\eeqa
We recover in this case the results of Pujol \cite{pujol}.\\
 We are looking for 
the fixed points by taking $ {dg_{12}\over d\log l}= {d\D_{ij}\over d\log l}=0$. 
For the system (\ref{rg2}), there are three fixed points
\beq
\label{qfixed1}
g=0~~~;~~~\Delta =0
\eeq
the trivial one, and
\beq
\label{qfixed3}
g=0~~~;~~~\Delta = {\epsilon \over 2} + {\epsilon^2 \over 4}
+ O(\epsilon^3)
\eeq
the fixed point corresponding to a decoupling of the models and
\beq
\label{qfixed4}
g={\epsilon^2 \over 4} + O(\epsilon^3)~~~;
~~~\Delta = {\eps+\eps^2\over 2}+ O(\epsilon^3)
\eeq
a new non-trivial one mixing both models.\\
 When we consider one Ising model coupled to a $3$ or $4$-state Potts model, we 
take the limit $\eps_1\to 0$ and  $\eps_2=\eps$, we find the following fixed 
points
\beqa
\label{ptfix1}
g^*_{12}=0&;&\D^*_{11}=\D^*_{22}=\D^*_{12}=0\\
\label{ptfix2}
g^*_{12}=0;~\D^*_{11}=0;~\D^*_{12}&=&{\eps\over 2}+{\eps^2\over 
4}+o(\eps^2);~\D^*_{12}=0+o(\eps)\\
\label{ptfix3}
g^*_{12}=0;~\D^*_{11}=0;~\D^*_{12}&=&{\eps\over 2}+{\eps^2\over 
4}+o(\eps^2);~\D^*_{12}={\eps\over 2}+o(\eps)
\eeqa
We have also three fixed points, the trivial one (\ref{ptfix1}), the one 
corresponding to a perfect decoupling of the disordered Ising and Potts models 
(\ref{ptfix2}) and a new fixed point
mixing {\it a priori} both models. Let us notice that $\D^*_{12}$ is 
undetermined at one loop and two loop calculations just enable to compute the 
first order in $\eps$. 
And finally, when one considers a $3-$state coupled to a $4-$ state Potts model, 
we only find the fixed points (\ref{ptfix1}), (\ref{ptfix2}).\\
In order to study the stability of each of these fixed points, we re-express the 
systems  (\ref{rg}) around the above solutions. Therefore, we write 
$g_{12}=g^*_{12}+\delta g_{12};~\D_{ij}=\D_{ij}^*+\delta \D_{ij}$ and keep only 
the smallest order in $\eps$. We thus obtain a linear system $\delta \dot{X}=AX$ 
with $X=(\delta\D_{11},\delta\D_{22},\delta\D_{12},\delta g_{12})$. All the 
information concerning the stability is contained in the  matrix $A$. For the 
R.G. equations (\ref{rg2}), $A$ is only a $2\times 2 $ matrix. By calculating 
the eigenvalues of $A$ for each fixed points, we can  show that the fixed points 
(\ref{qfixed1}), (\ref{qfixed3}) are
unstable. For the fixed point (\ref{qfixed4}), a numerical study of the flow 
(fourth order are needed for an analytical study of the stability of this fixed 
point!) shows that it is also unstable contrary to what was stated in 
\cite{pujol}. Consequently, the system is driven  in a strong coupling regime. 
Note that this result is surprising because it is in contradiction with Imry and 
Wortis arguments \cite{hui}. Nevertheless, we must not forget that we have done  
 a perturbative analysis at two loops, so another non-perturbative fixed point 
can not be ruled out (or a perturbative one which needs three loops terms to be 
stabilized). Therefore, it would be very interesting to study the critical 
behavior of this model numerically. \\
A similar analysis can be repeated for $\eps_1=0$. Thus, we can see that 
(\ref{ptfix1}) is unstable, that
(\ref{ptfix2}) is stable, and finally that (\ref{ptfix3}) is stable only
 in the space $g_{12}=0$. In fact, the fixed point  (\ref{ptfix3}) appears as a 
generalisation of  (\ref{qfixed3}) for different minimal models. We have 
represented in Figure 1, the projection on the flow in the 
$(g_{12},\Delta_{12})$ plane for two different initial conditions (points $A_0$ 
and $B_0$). We clearly see that the flow first try to
go away and then is driven by disorder at the origin corresponding to a perfect 
decoupling of the models. As it has been already noticed in \cite{cardy}, such a 
flow is unusual because it violates the $c-$theorem \cite{zamolod}. 
There are no corrections to the critical exponents which remain those of the 
corresponding decoupled disordered models. Note that in \cite{pujol}, a new
stable fixed point was found (for $N> 2$ $3-$state Potts models), but with 
similar
critical exponents that one disordered Potts model.\\

In this paper, we have analysed the behavior of two different coupled minimal 
models in presence
of disorder. This constitutes a new example of first order transition driven by 
randomness 
in a second order one (see Figure 1). Moreover, we have shown than the models 
factorize,
namely decouple. If we compare the critical exponent $\beta/\nu \sim 0.142$ 
found in \cite{cardy1}
for the disordered $6-$state potts model, it appears clearly different from the 
one
of the Ising or disordered $3-$state Potts model. Therefore,  there are no 
relations at a perturbative level between both models despite their central 
charges are similar. On the other hand, it would be very interesting to test 
numerically the behavior
of two  coupled $q_1,q_2$ Potts models with disorder and to analyse the 
non-perturbative area in relation with the $q_1\times q_2$-state Potts model. Indeed,
for large $g_{12}$ , we could imagine a cross-over phenomena in the system of 
two coupled
minimal models. The case of two similar Potts models is special because it seems 
to violate the Imry-Wortis argument perturbatively. It deserves a more accurate 
analysis, either analytically from, for example, the integrable point of view  
or numerically. Finally, such an analysis can be extended to the case of $N$ 
Ising models to $M$ Potts models. We expect similar behaviors except, maybe, for 
the cases $N$ or $M=2$.

\eject
{\bf Acknowledgements}\\\noindent
I would like to thank Vl. S. Dotsenko for helpful suggestions and stimulating 
discussions. I also acknowledge M. Picco and P. Pujol for useful discussions.

\eject
\begin{center}
{\bf FIGURE CAPTION}
\end{center}

\vskip 0.5 truecm
FIG.1 : The projection of two flows in the $(g_{12},\Delta_{12})$ plane for one 
Ising model coupled to a $3-$state Potts models with disorder.  $A_0,~B_0$ 
correspond to two different initial conditions. We clearly see that the flow is 
driven
by randomness at the origin. It corresponds to a decoupling regime.

\psfig{figure=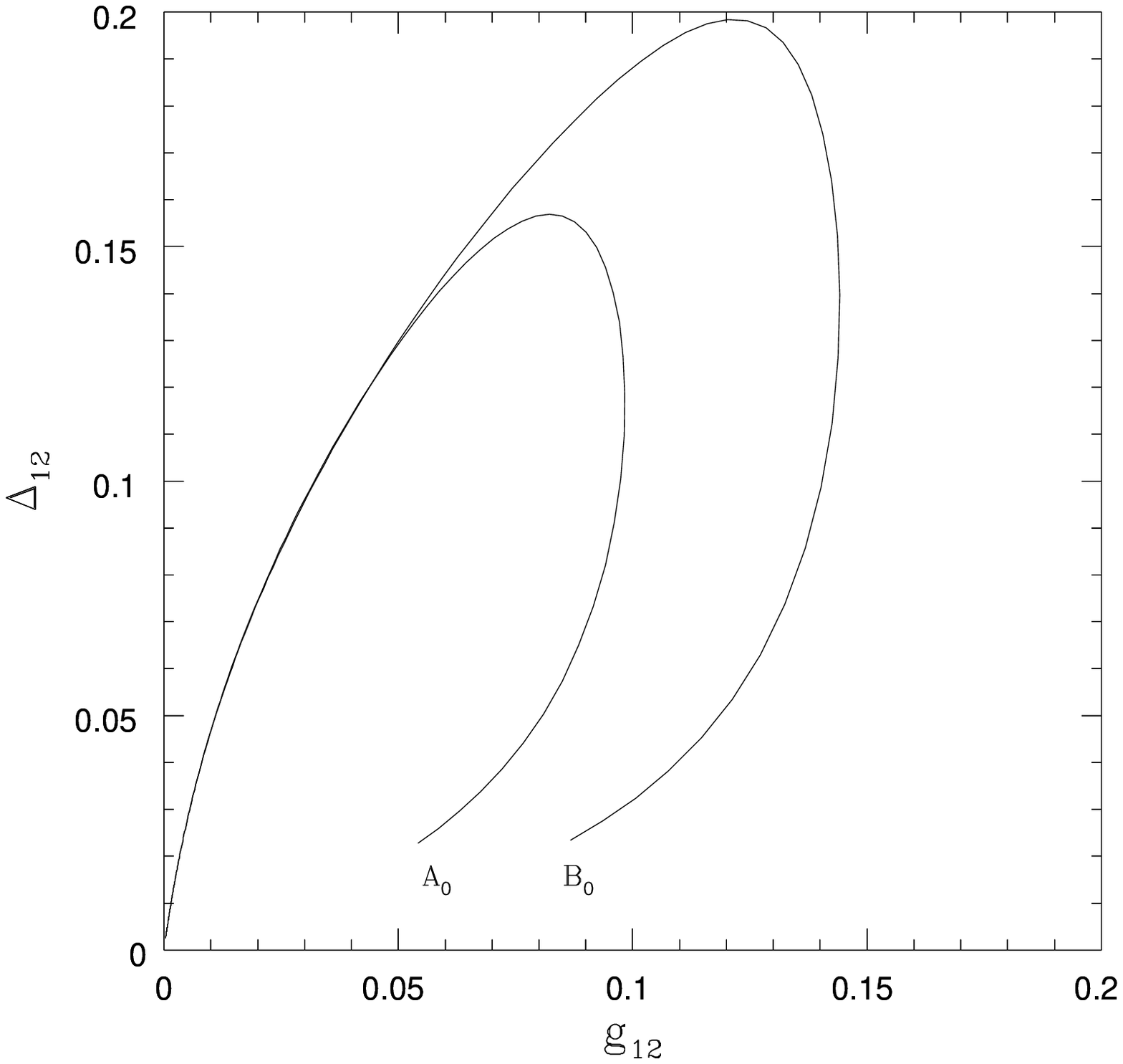,height=14cm,width=18cm}


\vskip .5 in
\baselineskip=1.6pt
\begin{thebibliography}{99}
%
\bibitem{ludwig}
A. W. W. Ludwig, Nucl. Phys. {\bf B285}, 97 (1987); Nucl. Phys.  {\bf B330}, 639 
(1990).
%
\bibitem{dots}
Vl. Dotsenko, M. Picco and P. Pujol, Phys. Lett {\bf B347}, 113 (1995); Nucl. 
Phys. {\bf B 455}, 701 (1995).
%
\bibitem{num}
E. Domany and S. Wiseman, Phys. Rev. {\bf E51}, 3074 (1995); Phys. Rev. {\bf 
E52}, 3469 (1995);
M. Picco, Phys. Rev. {\bf B54}, 14930 (1996).
%
\bibitem{harris}
A. B. Harris, J. Phys. {\bf A7}, 1671 (1974).
%
\bibitem{hui}
Y. Imry and M. Wortis, Phys. Rev. {\bf B19}, 3581 (1979);
K. Hui and N. Berker, Phys. Rev. Lett. {\bf 62}, 2507 (1989).
%
\bibitem{aiz}
M. Aizenman and J. Wehr, Phys. Rev. Lett. {\bf 62}, 2503 (1989).
%
\bibitem{cardy}
J. L. Cardy, J. Phys. {\bf A26}, 1897 (1996).
%
\bibitem{pujol}
P. Pujol, Euro. Phys. Lett. {\bf 35}, 283 (1996).
%
\bibitem{chen}
S. Chen, A.M. Ferrenberg and D. P. Landau, Phys. Rev. Lett. {\bf 69}, 1213 
(1992); Phys. Rev. {\bf E52}, 1377 (1995).
%
\bibitem{cardy1}
J. L. Cardy and J. L. Jacobsen, cond-mat/9705038.
%
\bibitem{picco}
M. Picco,  cond-mat/9704221 and private communication. 
%
\bibitem{dots1} 
Vl. S. Dotsenko and V.A. Fateev, Nucl. Phys. {\bf B240}, 312 (1984), {\bf 
B251}, 691 (1985).
%
\bibitem{vays}
L. Vaysburd, Nucl. Phys. {\bf B446}, 387 (1995).
%
\bibitem{simon}
P. Simon,  cond-mat/9705292, to be published in Phys. Lett. B.
%
\bibitem{simon1}
P. Simon, in preparation
%
\bibitem{dots2}
Vl. S. Dotsenko {\it et al.}, in preparation.
%
\bibitem{zamolod}
A. B. Zamolodchikov, Sov. J. Nucl. Phys. {\bf 46}, 6 (1987); Advances Studies in 
Pure
Mathematics {\bf 19}, 641 (1989). 
\end{thebibliography}
\end{document}